\documentclass[12pt]{article}
\usepackage{epsfig}
\usepackage{fullpage}
\psfigdriver{dvips}

\input{epsf}
\begin{document}

\def\C{{\rm\kern.24em \vrule width.02em height1.4ex
depth-.05ex\kern-.26em C}}
\def\N{{\rm I\hspace{-0.4ex}N}}
\def\R{{\rm I\hspace{-0.4ex}R}}
\def\rit{{\rm I\hspace{-0.4ex}R}}
\def\grad{\hbox{ grad }}
\def\curl{\hbox{curl }}
\def\div{\hbox{ div }}
\def\supp{{\rm supp}\:}
\def\Lip{{\rm Lip}}
\def\sign{{\rm sign}\:}
\def\dist{{\rm dist}}
\def\diam{{\rm diam}\:}
\def\const{{\rm const.}\:}
\def\meas{{\rm meas}\:}
\def\oOmega{\overline{\Omega}}
\def\ep{\varepsilon}
\def\be{\begin{equation}}
\def\ee{\end{equation}}
\def\beq{\begin{equation}}
\def\eeq{\end{equation}}
\def\ds{\displaystyle}
\def\ts{\textstyle}
\def\qed{\rule{1ex}{1ex}}

\title{\bf  The bifurcation diagrams for the Ginzburg-Landau system for 
superconductivity}
\author{Amandine Aftalion\footnote{Laboratoire d'analyse num\'erique,
B.C.187,
Universit\'e Pierre et Marie Curie,
4 Place Jussieu,
75252 Paris cedex 05, France. }$\ $
and Qiang Du\footnote{Department of Mathematics, 
Hong Kong University of Science and Technology,
Clear Water Bay,
Kowloon, Hong Kong.}}
\maketitle

\abstract{
In this paper, we provide the different types of bifurcation 
 diagrams for a superconducting cylinder placed in a magnetic field 
 along the direction of the axis of the cylinder. The computation is based
 on the numerical solutions of the
 Ginzburg-Landau model by the finite element method. The response of the material depends on the values of the exterior field, the Ginzburg-Landau parameter and the size of the domain.
 The solution branches in the different regions of the bifurcation diagrams 
are analyzed and open mathematical problems are mentioned. 
}

\section{Introduction}
In this paper, we want to give a detailed description of the bifurcation
diagrams of 
an infinite superconducting cylinder of cross section $\Omega$, submitted 
to an exterior magnetic field $h_0$.  The response of the material varies 
greatly according to the value of $h_0$, the size of the 
cross section and the  Ginzburg-Landau parameter $\kappa$
that characterizes the material:
superconductivity appears in the volume of the sample for low fields 
and small samples, under the form of vortices for higher fields, 
bigger samples and larger values of $\kappa$, and is destroyed
for high fields. 
The type of response of a superconducting material has been studied 
numerically and theoretically by various authors in various asymptotic 
regimes \cite{AD,CDGP,DPS,dPFS,DGP,DGP2,GP,SS1,S1,S2}. Here, we want 
to give a complete numerical picture of the bifurcation diagrams 
for all values of the parameters. This type of computation has already 
been done in dimension 1 by one of the authors using auto \cite{AT1}. 
Here, the computation is made in a 2 dimensional domain using
numerical solutions of the well-known Ginzburg-Landau model \cite{T}, based on
a code first developed in \cite{DGP}. We examine the behavior of the
energy, the magnitude of the order parameter and the magnetization versus the magnetic field for various solution branches.
We also provide some analysis on the detailed findings such as the stability
of solutions.

The paper is organized as follows: the Ginzburg-Landau model is briefly
stated in Section 2 and the main features of the numerical codes used 
in the computation are described Section 3. The complete phase diagrams
are given in Section 4, along with detailed analysis.
A conclusion is given in Section 5.

\section{The  Ginzburg-Landau model}
Let $\Omega$ denote 
a two dimensional bounded domain which represents the cross section
of a three dimensional cylinder occupied by the
superconducting sample. Assume that the cylinder is homogeneous along
its axis and a constant applied field  $H_0$ is placed along the axis direction
as well. Then, the Gibbs free energy ${\cal G}$ may be written
in the following form \cite{T}:
$$
\begin{array}{l}
   {\cal G}(\psi,A) = \int_{\Omega} 
\left(f_n + \alpha |\psi|^2
+ {\beta\over 2} |\psi|^4 \right) d\Omega \\
\qquad +
\int_{\Omega}  \left[{1\over{2m_s}}\left|\left(i\hbar\nabla
+ {{e_s A }\over c}\right)\psi\right|^2
 + {{|\mbox{curl } A|^2}\over{8\pi}} - {{\hbox{curl } A
\cdot H_0}\over{4\pi}}\right]
 d\Omega \, .
\end{array}
$$
Here,  $\psi$ is the (complex-valued) order parameter, $A$ is
the magnetic potential, $\hbox{curl } A$ is the magnetic field,
$\alpha$ and $\beta$ are constants (with respect to the  space
variable $x$) whose values depend on the temperature, $c$ is the
speed of light, $e_s$ and $m_s$ are the charge and mass,
respectively, of the superconducting charge-carriers, and
$2\pi\hbar$ is Planck's constant. 

After proper nondimensionalization, we can reformulate the free energy
functional as:
$$
   {\cal G}(\psi,A) = \int_{\Omega} \left|\left(\nabla - i
 A \right)\psi\right|^2
 +
{\kappa^2 \over 2} \left(1- |\psi|^2\right)^2
+ |\mbox{curl } A - h_0|^2
 d\Omega , 
$$
where $\kappa$ is the Ginzburg-Landau parameter representing the ratio of
the penetration depth and the coherence length, $h_0$ the 
applied magnetic field and $d$ the characteristic 
 size of the domain $\Omega$, that is $\Omega=d D$ where $D$ is a fixed domain.

The system 
that we are going to study is the following Ginzburg-Landau equations
derived as the Euler-Lagrange equations for the minimizers of the functional
${\cal G}$ \cite{GL}:
\begin{equation}
\label{GL}
\left\{ 
\begin{array}{l}
 -(\nabla-iA) ^2 \psi  = \kappa^2
\psi(1-|\psi|^2)  \hbox{ in } \Omega, \\ 
-\hbox{curl curl } A=|\psi|^2 A+\frac{i}{2}(\psi^* \nabla \psi -\psi
\nabla \psi^*)
\hbox{ in } \Omega ,
\end{array} 
\right.
\end{equation}
which are supplemented by the boundary conditions 
$$\left\{ 
\begin{array}{l}
(\nabla \psi - i A \psi) \cdot \hbox{n} =  0   \hbox{ on }
\ \partial\Omega, \\
 \hbox{curl } A  =  h_0   \hbox{ on }
 \ \partial\Omega, 
\end{array} \right.
$$
and gauge constraints
$$\left\{ 
\begin{array}{l}
\hbox{div } A= 0 \hbox{ in } \Omega, \\ 
A\cdot \hbox{n}   =  0   \hbox{ on }
 \ \partial\Omega .
\end{array} \right.
$$
Here,  $\partial \Omega$ is the boundary of $\Omega$ and $\hbox{n}$ is its 
unit outer normal. With the above nondimensionalization, 
$|\psi|$ takes values between $0$ and $1$: the normal state corresponds to
 $|\psi|=0$ while 
the Messiner state corresponds to 
$|\psi|=1$.

This Ginzburg-Landau model has a special family of solutions called 
the {\em normal solutions}: $\psi=0$ and $\hbox{curl } A=h_0$, which 
correspond to the situation where superconductivity is destroyed. According to 
the values of the different parameters $\kappa$, $d$ and $h_0$, 
the system may have other solutions: {\em superconducting 
solutions}, for which $\psi$ is never 0 and {\em vortex solutions} 
for which $\psi$ has isolated zeroes (see Figure \ref{fig.01} and \ref{fig.02}). For a complete introduction 
to the topic, one may refer to \cite{T}.

\hfill

\section{Numerical codes}
The numerical code that we are using is based on the finite element 
approximations of the Ginzburg-Landau model, first proposed  in 
 \cite{DGP} and later  used in many settings, see \cite{DGP2,wang,wang2} 
for instance. The codes have many different variants that can be
used to simulate the solutions of the Ginzburg-Landau models in
the high-$\kappa$ setting \cite{CDGP,dugray}, 
thin films with variable thickness \cite{CDG},
samples with normal inclusions \cite{DGP2} as well as layered materials.
Here, we choose the standard version that solves the
Ginzburg-Landau equations on a rectangular domain $\Omega$ and $d$ will 
denote the characteristic size of the rectangle. 
We note that other numerical methods,
such as the gauge invariant difference methods \cite{du}, 
i.e, the so-called bond-and-link variable methods or the method
of eigen-functions, have also been used to compute the
phase diagrams for the Ginzburg-Landau equations \cite{BPV,dspg}.

In our implementation of the finite element approximation, 
we use a uniform triangular
grid with piecewise quadratic polynomials for both $\psi$ and $A$.
It is shown that (see \cite{DGP} for instance), if $(\psi_\delta,A_\delta)$ 
is the finite element solution on a given mesh with mesh size $\delta$,
the convergence of the approximation is assured, and
asymptotically, we have 
$$\|\psi-\psi_\delta\|_2 + \|A-A_\delta\|_2 = O(\delta^3 )\;$$
where $\|\cdot\|_2$ denote the standard mean square $L^2$ norm.
For each set of calculation, we refine the mesh size until numerical
convergence is evident.

\begin{figure}[htb]
\vspace{1.in}
\centerline{
\epsfxsize=1.2in\epsfbox{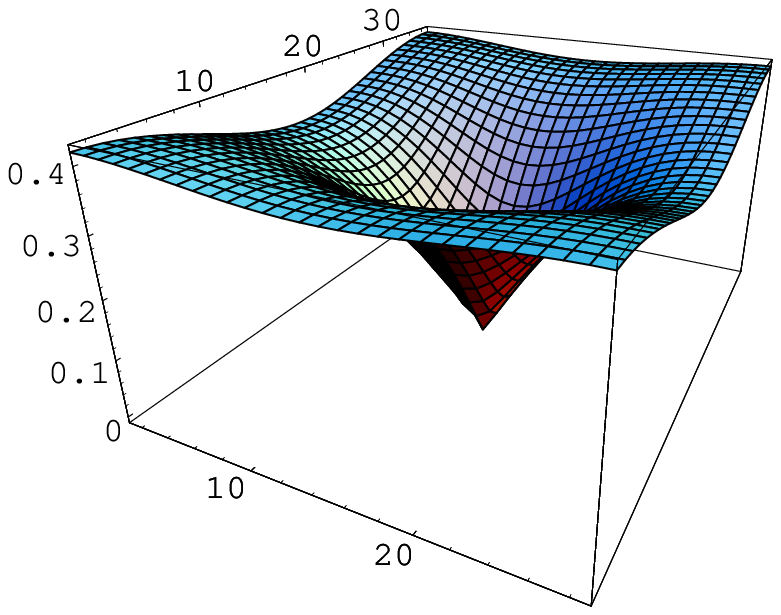}$\qquad\quad$
\epsfxsize=1.2in\epsfbox{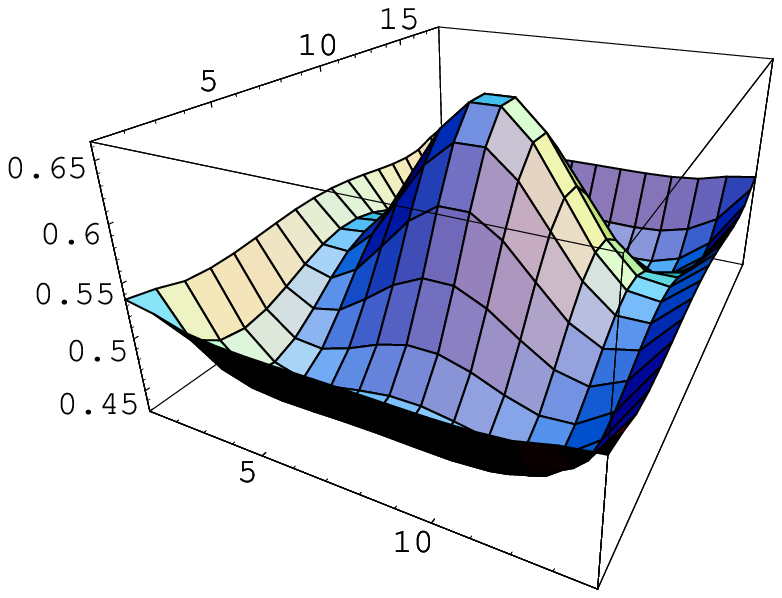}$\;\qquad\;$
\epsfxsize=0.7in\epsfbox{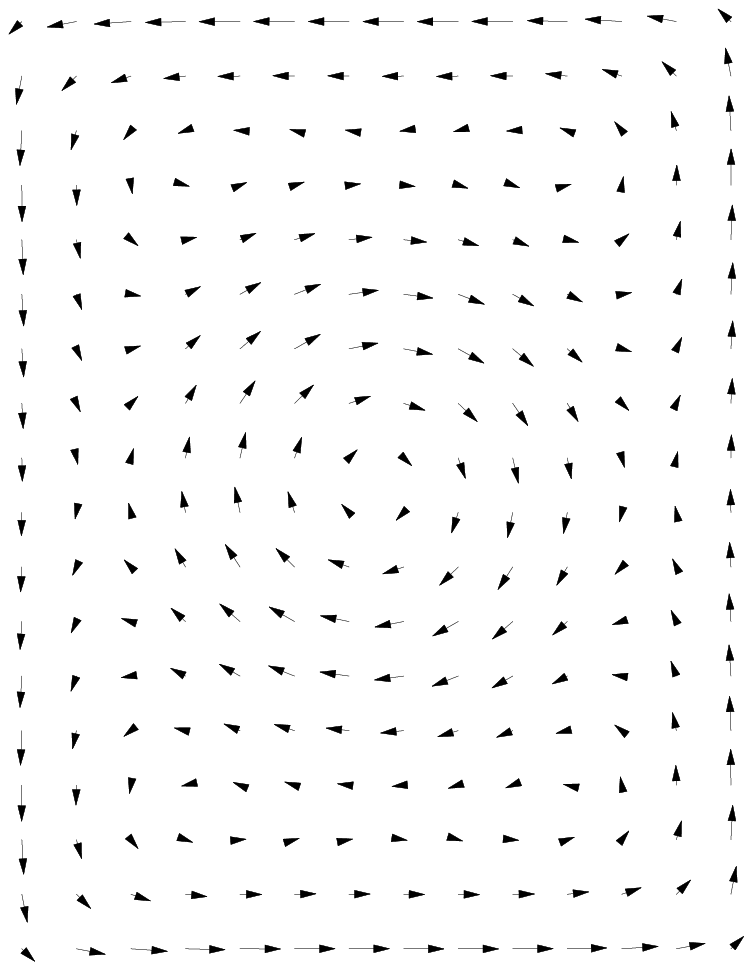}
}
\vspace{-1.5cm}
\caption{A single vortex solution of (\ref{GL}).
}\label{fig.01}
\end{figure}

\begin{figure}[htb]
\vspace{0.8in}
\centerline{
\epsfxsize=1.2in\epsfbox{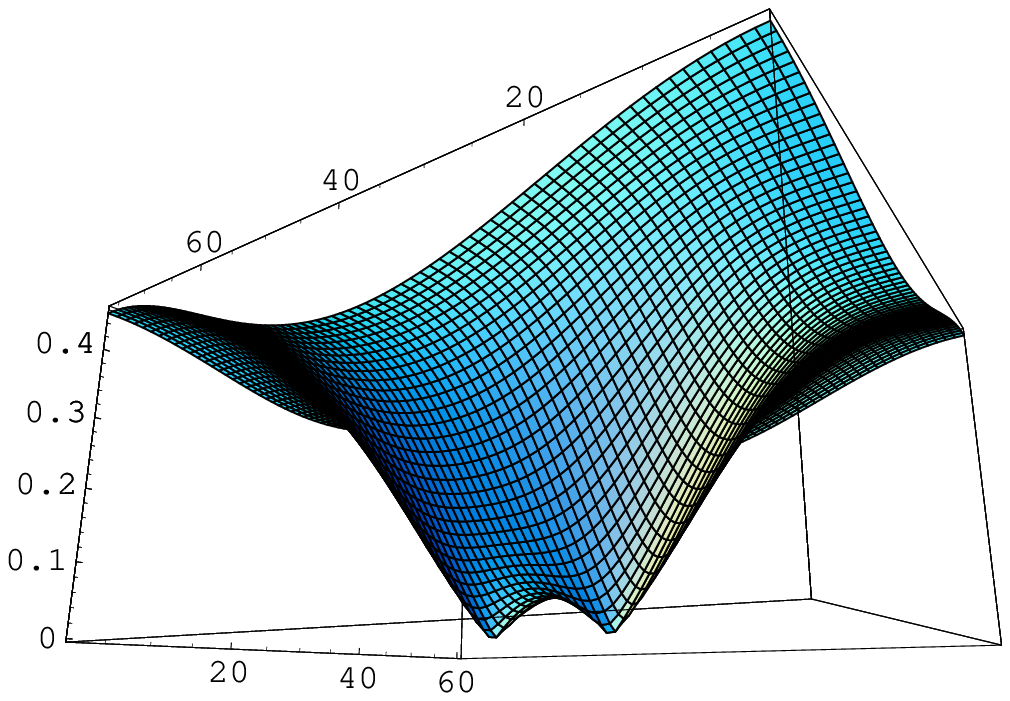}$\quad\qquad$
\epsfxsize=1.2in\epsfbox{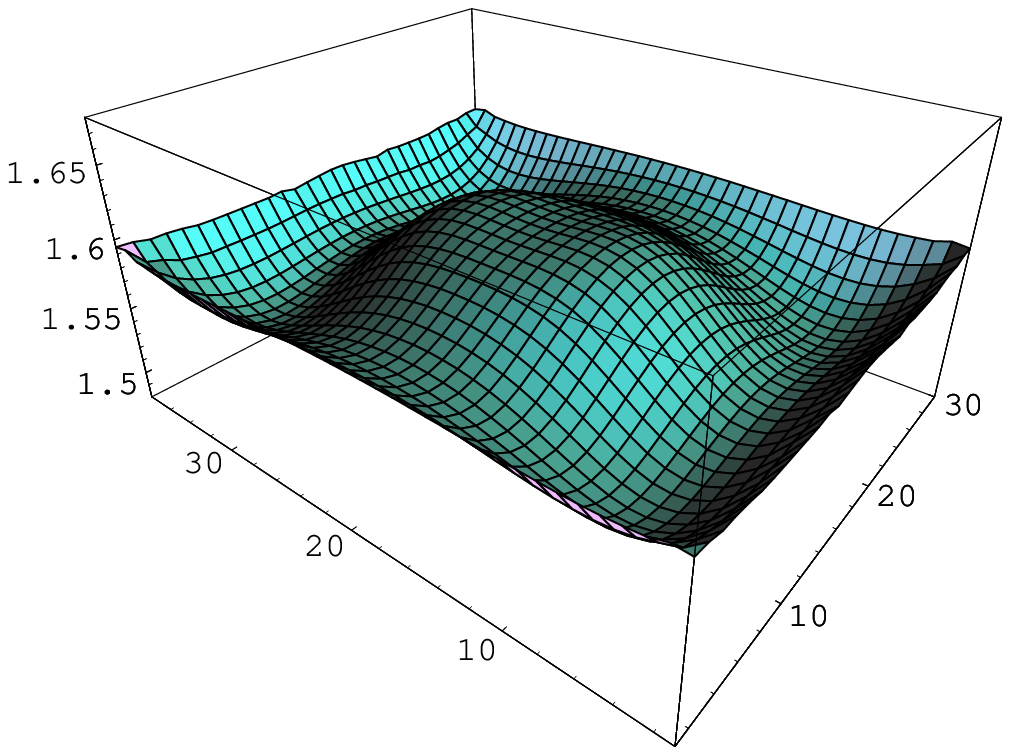}$\;\qquad\;$
\epsfxsize=0.7in\epsfbox{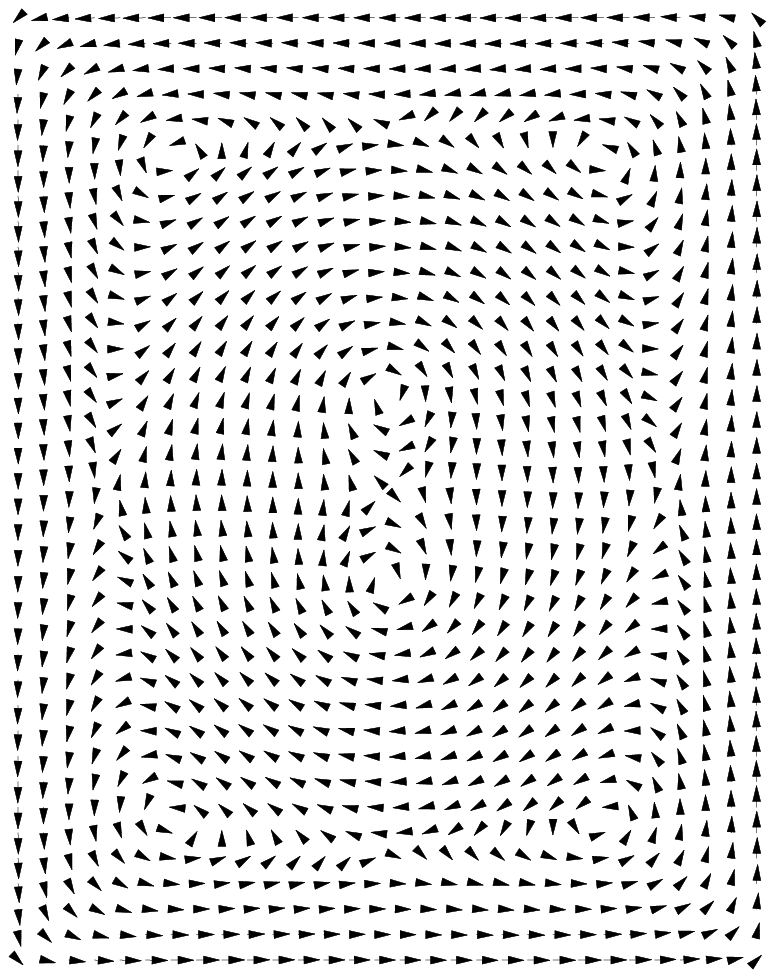}
}
\vspace{-1.5cm}
\caption{A solution of (\ref{GL}) with two vortices.}\label{fig.02}
\end{figure}

In Figures \ref{fig.01} and \ref{fig.02}, 
we present a few typical plots for the
numerical solutions of the equation (\ref{GL}).
For each solution, the plots include a surface plot
of the magnitude of the order parameter, a surface plot of the
magnetic field given by ${\rm curl}\, A$ and a vector plot of the
superconducting current. 
In  Figure \ref{fig.01},  we have a solution with a single vortex 
at the center of the domain which corresponds to the parameter values
$\kappa=0.23$,  $d=16.8$ and $h_0=0.563$. 

In  Figure \ref{fig.02}, we present the plots for
a solution with two vortices corresponding to $\kappa=0.8$,  
$d=4$ and $h_0=1.2$.

For fixed $\kappa$, $d$ and $h_0$, we are interested in 
finding the number of solutions of (\ref{GL}) and their stability. A continuation in the parameter spaces is used for getting solutions with
different parameter values.
With $\kappa$, $d$ given but $h_0$ allowed to vary, 
for a computed solution branch, we plot 
$\|\psi\|_\infty$, the maximum magnitude of the order 
parameter $\psi$ in the domain, the free energy $\cal G$ and  
the magnetization versus the applied field $h_0$. 
These phase diagrams or bifurcation diagrams will give us  
information on the solutions (number and stability) for each $h_0$. 
They were drawn by Ginzburg \cite{G} in some limiting cases
 of the parameters. Here we want to give a more
complete description of these diagrams for all values of the parameters.

The results of our numerical computations allow us to separate 
the $\kappa$-$d$ plane into different regions depending on the 
shape of the bifurcation diagram. A solution branch may 
exist in one region but may cease to exist in
another one,  it may also develop hysteresis in some regions.

\hfill

\section{The bifurcation diagrams}

\begin{figure}[htb]
\begin{center}
\input{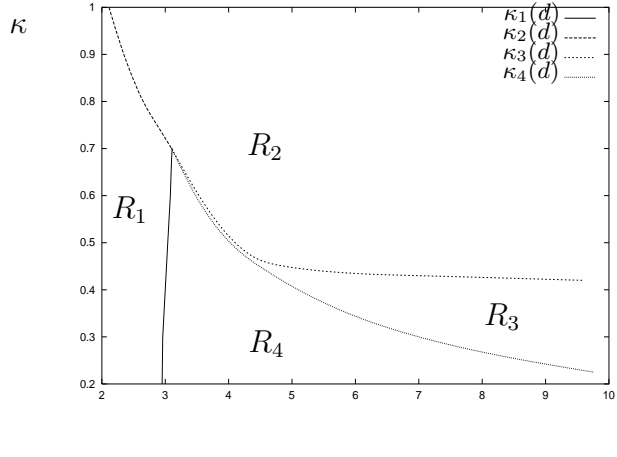}
\end{center}
\vspace{-0.5cm}
\caption{The curves $\kappa_1(d)$, 
$\kappa_2(d)$, $\kappa_3(d)$ and $\kappa_4(d)$}
\label{curves}
\end{figure}

It is well known that for large fields ($h>h_*$), the only solution is the 
normal solution \cite{GP}. For smaller fields ($h<h_*$), the normal solution 
always exists  but there are other solutions which display four different 
types of behaviors.
These behaviors depend on the values of $\kappa$ and $d$. 
In Figure 
\ref{curves}, we have plotted four curve segments $\{\kappa_i(d)\}_{1}^{4}$
 separating the $\kappa-d$ plane into four regions $\{R_i\}_{1}^4$.

The four curve segments in Figure \ref{curves} 
 are the results of the computation described in the 
earlier section.
All four curves meet close to $\kappa=1/\sqrt2$, $\kappa_2(d)$ is of the 
form $2.112/d$, $\kappa_3(d)$ is tending to 0.4 at infinity.
The four regions correspond to the four types of behaviors 
for the bifurcation diagrams. For convenience, 
for each $i=1,2,3,4$, we use $d=d_i(\kappa)$ to denote
the inverse function of the function $\kappa=\kappa_i(d)$ wherever the 
inverse is well-defined.
 What distinguishes the different regions
 are features like the existence 
(or the lack of existence) of
vortex solutions, the global and local stability of solutions, and the
hysteresis phenomena.


We now provide detailed descriptions
of the solution behavior for each region in Figure \ref{curves}.

\hfill

\noindent
{\bf Region 1}: $d<d_1(\kappa)$ and 
$d <d_2(\kappa)$.

\begin{figure}[ht]
\begin{center}
\input{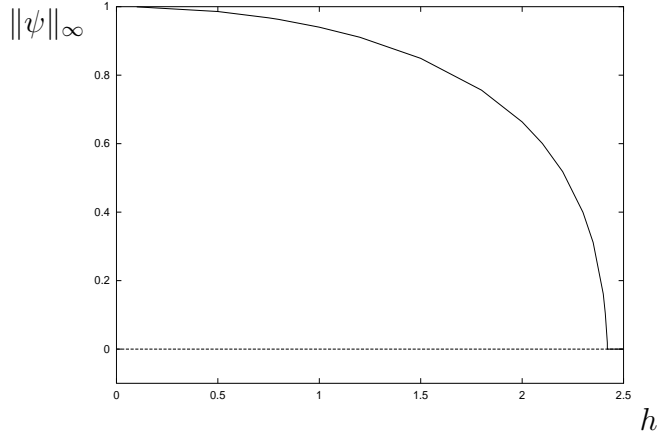}
\end{center}
\vspace{-0.5cm}
\caption{The bifurcation curve for $d=2.0$ $\kappa=0.3$}
\label{ux.6}
\end{figure}

\begin{figure}[ht]
\begin{center} 
\input{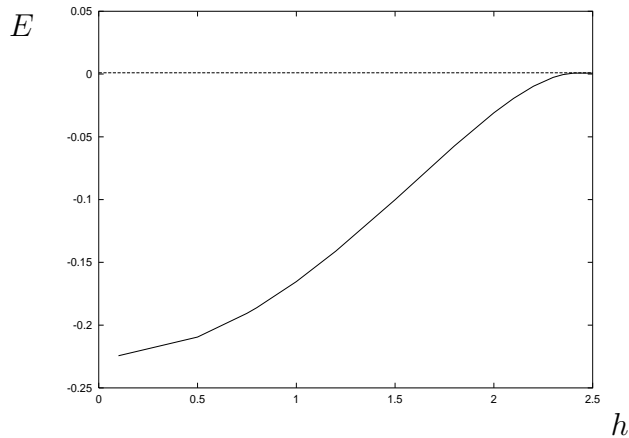}
\end{center}
\vspace{-0.5cm}
\caption{The energy for $d=2.0$ $\kappa=0.3$}\label{ex.6}
\end{figure}

\begin{figure}[ht]
\begin{center} 
\input{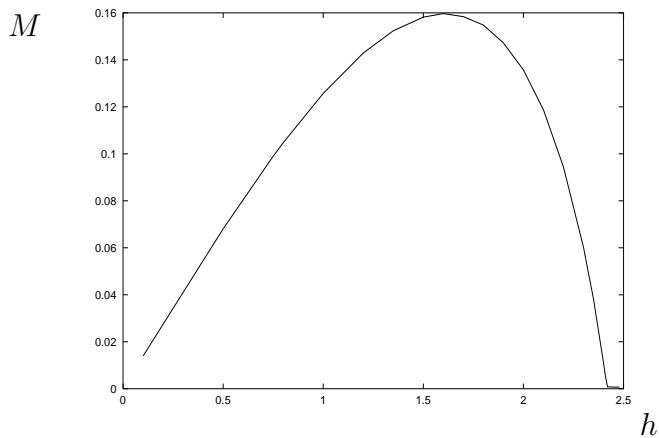}
\end{center}
\vspace{-0.5cm}
\caption{The magnetization for $d=2.0$ $\kappa=0.3$}\label{mx.6}
\end{figure}

\noindent
This corresponds to the situation where the cross 
section of the superconducting sample is small enough. 
The bifurcation diagram 
is illustrated in Figure \ref{ux.6}. The corresponding plot of
the energy is given  in Figure \ref{ex.6} and the magnetization curve
in Figure \ref{mx.6}.

Throughout this region, there is a unique non-normal solution for $h<h_*$. 
This solution is a superconducting solution
which is the global minimizer of the free energy ${\cal G}$.
The curve $|\psi|_\infty$ against $h$ is monotonically decreasing. 
 When increasing the field, the magnitude of the superconducting solution decreases until it turns normal at $h_0=h_*$. Conversely, when decreasing the field, the normal solution turns superconducting also for $h_0=h_*$.
The transition to the normal solution is 
of second order, that is the energy of the superconducting solution tends 
to the energy of the normal solution at the transition and there is no 
hysteresis phenomenon. This can also be seen on the magnetization curve 
of Figure \ref{mx.6}. 

There is no vortex solution for the parameters $(d,\kappa)$ in this region.
This reflects the fact that  $d$ is too small to
allow enough room for a vortex to exist since a vortex core
is of typical size $C/\kappa$. 

In \cite{AD}, it is proved that for 
$d< \min\{ d_0, d_1 /\kappa\}$,
 a solution of the Ginzburg-Landau equations has no vortex, and
for such  a solution, $\psi$ is almost 
constant in the domain and is unique up to multiplication by a constant of modulus 1. 
Also it is proved that the bifurcation curve for $|\psi|_\infty$ 
is decreasing and that $h_*$ is asymptotically $C/d$ when $d$ tends to 0. More precisely, for a disc, when $d$ tends to 0, the limiting curve is $\|\psi\|_\infty^2=1-d^2h_0^2/2\kappa^2$.  

\hfill

\noindent
{\bf Region 2:} $\kappa>\kappa_2(d)$ and $\kappa>\kappa_3(d)$ . 

\begin{figure}[ht]
\begin{center}
\input{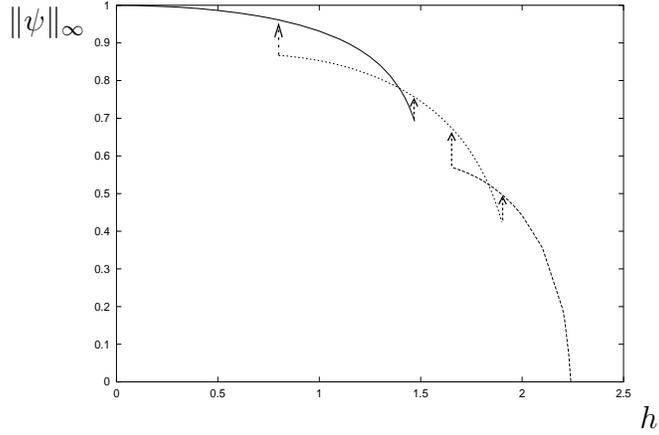}
\end{center}
\vspace{-0.5cm}
\caption{The bifurcation curve for $d=3.2$ $\kappa=1$}\label{ux3.2}
\end{figure}

\begin{figure}[ht]
\begin{center} 
\input{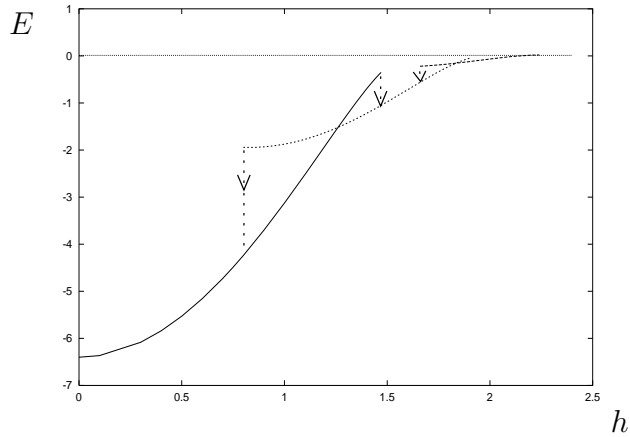}
\end{center}
\vspace{-0.5cm}
\caption{The energy for $d=3.2$ $\kappa=1$}\label{ex3.2}
\end{figure}

\begin{figure}[ht]
\begin{center} 
\input{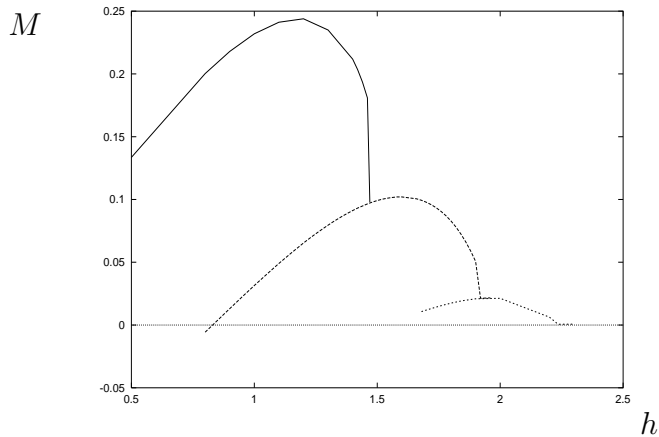}
\end{center}
\vspace{-0.5cm}
\caption{The magnetization for $d=3.2$ $\kappa=1$}\label{mx3.2}
\end{figure}

\noindent
In contrast with Region 1, this region corresponds to the situation where the typical size of vortices ($C/\kappa$) is small enough compared to the size of the domain.
This region displays the typical type II behavior of superconductors.
The bifurcation diagram is illustrated in Figure \ref{ux3.2},
the energy in Figure \ref{ex3.2} and the magnetization curve is given
in Figure \ref{mx3.2}.

It has been well understood both physically
and in more recent years mathematically that, 
for sufficiently large $\kappa$, there are 
vortex solutions which are the global minimizers of the free energy
 for a certain range of fields. The number of vortices depends  on the strength of the applied field. 
The maximum number of vortices 
increases with $d$ and $\kappa$, a type-II behavior. 

For very small fields, 
the global minimizer is the superconducting solution (solid line). As the 
field is increased, the superconducting solution loses its global stability 
($h=h_{c_1}$ and for even larger fields loses its local stability, 
 $h=h_0^*$). Then  the global minimizer starts to nucleate vortices.
In Figures \ref{ux3.2},  \ref{ex3.2}, \ref{mx3.2}, the solution branch 
corresponding to a one vortex solution is illustrated by a dotted line 
and to a two vortex solution by a dashed line.

Let $n$ be a non-negative integer (the flux quanta).
Hysteresis phenomena occur in our experiments 
when going from a solution with $n$ 
vortices to a solution with $n+1$ vortices: when increasing the field, 
the solution with $n$ vortices remains locally stable, though it is no 
longer a global minimizer until $h_n^*$ where it jumps to a solution
with $n+1$ vortices. 
Similarly when decreasing the field, the solution with $n+1$ vortices 
remains locally stable down to $h^{n+1}_*< h_n^*$. 
Such a hysteresis 
pattern is typical for parameters in this region. 
The fields $h_n^*$ and $h^{n+1}_*$ correspond to the 
critical field where the vortex solution loses its local stability
and their values can be identified from Figures \ref{ux3.2} and
\ref{ex3.2} (transition from superconducting solution to 1 vortex solution and from one vortex to two vortices). 

Nevertheless, the transition from normal to the last vortex 
solution is of second order since the vortex solution is stable, as can 
be seen from the energy and magnetization curves. This transition occurs at a field that is usually called $h_{c_3}$ in the literature.

There have been many numerical works on these vortex solutions: see for instance \cite{DPS}.  Recently, a rigorous analysis of vortices, their number and the critical thermodynamic fields in the high kappa limit has been done by Sandier and Serfaty \cite{SS1,S1,S2}: in particular they obtain in the high kappa limit that $h_{c_1}$ (the field for which the energy of the superconducting solution is equal to the energy of the one vortex solution) is of order $C\log\kappa $ and the field $h_n^*$ is asymptotic to $h_{c_1}+n\log \log \kappa $. 

The onset 
 of superconductivity in decreasing fields (instability of normal solutions and computation of the fields of nucleation) has been analyzed by Bernoff and Sternberg \cite{BS} and del Pino, Felmer, Sternberg \cite{dPFS}. Their works provide, as $d$ and $\kappa$ tends to $\infty$,  an asymptotic development of $h_{c3}$, the field at which the normal solution bifurcates to a vortex solution. This is what is called surface superconductivity.
A linearizing the Ginzburg-Landau equation has been done near the normal solution. In the high kappa limit, their computation yields
\beq\label{highk}
h_{c_3} \sim {\kappa^2 \over \lambda_1} +{C\kappa \kappa_{max}}+o(\kappa)\eeq
where $\kappa_{max}$ is the maximal curvature of the domain and $\lambda_1$ is the first eigenvalue of the linearized problem and is approximately equal to $0.59$.
In the high $d$ limit, it yields
\beq\label{highd}
h_{c_3} \sim {\kappa^2 \over \lambda_1} +{{C\kappa \kappa_{max}}\over {d^2}}+o({1\over {d^2}}).
\eeq
This expansion is consistent with the work of Saint James and de Gennes \cite{SJDG} who got the first term of this expansion in the case of an infinite plane in one dimension. In two dimensions, one has to take into account the curvature of the cross section. In the case of the disc, the equivalent of expansions (\ref{highk}) and (\ref{highd}) have been carried out by \cite{BPT} in the limit $\kappa d$ large.

The hysteresis phenomenon has been rigorously analyzed in \cite{LD} for the superconducting solution ($n=0$).

Some interesting questions to be addressed here are: 
what are the critical fields
for the vortex solutions losing their local
stability ($h_n^*$ and $h_*^n$)? can we find them in some asymptotic limit such as $d$ large or $\kappa$ large or $\kappa d$ large? and can we prove the existence of these hysteresis phenomena for the general case?

\hfill

\noindent
{\bf Curve $\kappa_2(d)$:}

\noindent
 For fixed $\kappa$ above 0.7, when $d$ is increased from 0, the point 
$(d,\kappa)$ is first in Region 1. Then
it reaches the critical value $d_2(\kappa)$. For $d=d_2(\kappa)$, the bifurcation diagram $\|\psi\|_\infty$ vs $h$ is decreasing and the superconducting solution bifurcates from the normal solution at $h=h_*$.
 For $d$ a little bigger than $d_2(\kappa)$, there is a vortex solution bifurcating from the normal solution close to $h_*$. Hence for $d=d_2(\kappa)$, at $h=h_*$,
   the linearized problem near the normal state has two eigenfunctions: one without vortices and one with a vortex.
 This is how uniqueness of solution is lost when increasing $d$, 
though the assertion needs to be proved mathematically.

Let $D$ be the fixed domain such that $\Omega = dD$. Then a bifurcated solution near the normal state $(0,h_0a_0)$ (where $a_0$ is such that $\curl a_0=1$ in $D$ and $a_0\cdot n =0$ on $\partial \Omega$) is of the form $(\ep\phi,h_0a_0+\ep B)$. Let $\omega=h_0 d$. The second variation of $\cal G$ near the normal state is
$${{\partial {\cal G}^2}\over {\partial \ep}^2}(\ep\phi,h_0a_0+\ep B)={1\over { d^2}}\int_D |(i\nabla+\omega^2a_0)\phi|^2-\kappa^2d^2|\phi|^2+ |\curl B|^2.$$
Let $$\lambda(\omega)=\inf \Bigr ( \int_D |(i\nabla+\omega^2a_0)\phi|^2, \ \|\phi\|_{L^2}=1, \ \phi\in H^1(D,\C )\Bigl ).$$
Hence if $\lambda(\omega)>\kappa^2d^2$, the normal solution is stable, if $\lambda(\omega)<\kappa^2d^2$, the normal solution is unstable and if $\lambda(\omega)=\kappa^2d^2$, degenerate stability occurs. For the eigenvalue $\lambda(\omega)=\kappa^2d^2$,  bifurcation of non normal solutions occurs. Thus, one has to study
\begin{equation}
\left\{
\begin{array}{l}
\label{lin}
(\nabla -i\omega^2a_0)^2\phi=\lambda (\omega) \phi \quad \hbox{in}\quad D,\\
{{\partial \phi}\over {\partial n}}=0 \quad \hbox{on}\quad \partial D,
\end{array}
\right.
\end{equation}
with $\lambda(\omega)=\kappa^2d^2$. For most values of $\kappa$ and $d$, the field $h$ such that the first eigenvalue $\lambda(\omega)$ is equal to $\kappa^2d^2$ yields a single eigenfunction. In region 1, the eigenfunction has no vortex while in region 2, the eigenfunction has a vortex. Thus, the curve $\kappa_2(d)$ corresponds to those values of $\kappa$ and $d$ for which the eigenvalue has two different eigenfunctions, one without vortices and one with a vortex. That is, 
on $\kappa_2(d)$,  the vortex state starts to exist. 

This situation has been studied in the case of a ball in \cite{BPT}. See also \cite{BA} for more recent developments. In this case, the solutions of (\ref{lin}) with $n$ vortices are of the form $\xi_n(r) exp(in\theta)$ and have eigenvalue $\lambda(\omega,n)$. In particular in \cite{BPT} they draw the function $\lambda(\omega,n)$ versus $\omega$. The curves $\lambda(\omega,0)$ and $\lambda(\omega,1)$ intersect exactly once for $\omega=\omega_0$ and $\lambda=\lambda_0$. Because of the bifurcation condition $\lambda(\omega)=\kappa^2d^2$, it implies that $\kappa^2d^2=\lambda_0$, hence the curve $\kappa_2(d)$ is of the form $\kappa d= constant$. It would be interesting to give a rigorous proof that the curves $\lambda(\omega,0)$ and $\lambda(\omega,1)$ intersect only once for the case of the disc and for the case of a more general domain. In region 1, that is below $\kappa_2(d)$, the first eigenfunction is simple and leads to a solution without vortex. In region 2, that is above $\kappa_2(d)$, the first eigenfunction is simple and leads to a solution with one vortex, but we expect that there is also an eigenfunction with no vortex for a lower field $h$.

Similarly, the curves $\lambda(\omega,n)$ and $\lambda(\omega,n+1)$ also intersect only once on the numerics of \cite{BPT} which means in our setting that there are curves $\kappa d=C_n$ at which the eigenvalue has two eigenfunctions with $n$ and $n+1$ vortices. Above $\kappa d=C_n$,  a solution with $n+1$ vortices starts to bifurcate from the normal state and below it, a solution with $n$ vortices starts to bifurcate from the normal state, so that the curve $\kappa d=C_n$ are the critical curves for the existence of $n+1$ vortices.

In the general case that we are studying, it is totally open to prove that there is a unique value of $\omega$ such that $\lambda(\omega)$ has two eigenvalues, one with a vortex and one without. This would yield to $\kappa_2(d)=C/d$, which is what we have found numerically. Moreover, we observe that the field of bifurcation $h_*$ satisfies $h_* d=\omega$ hence is constant along $\kappa_2(d)$. 

Taking this analysis of bifurcation a little further allows us to define
$${\cal H}(\kappa,d)=\{ h,\ s.t. \ \lambda(\sqrt{h d})= \kappa^2 d^2\}.$$
In region 1, we expect that ${\cal H}$ has a single element while in Region 2, we expect this set to have several elements corresponding to the various branches of solutions with several vortices bifurcating from the normal state. But this analysis is open even in the case of the disc.

\hfill
 
\noindent
{\bf Region 3: $\kappa_4(d)<\kappa<\kappa_3(d)$.} 

\begin{figure}[ht]
\begin{center}
\input{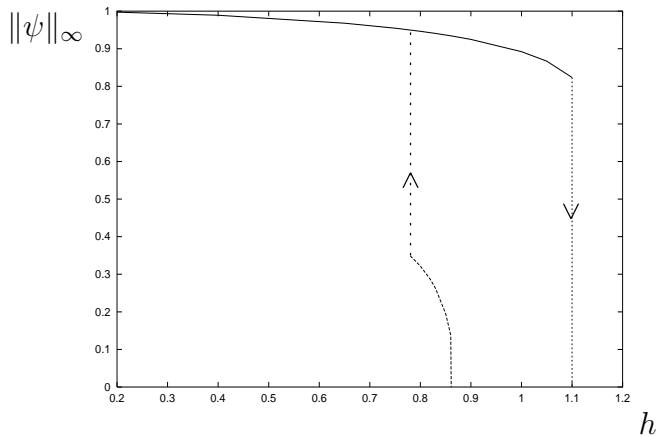}
\end{center}
\vspace{-0.5cm}
\caption{The bifurcation curve for $d=6.274$ $\kappa=0.35$}\label{ux2.196}
\end{figure}

\begin{figure}[ht]
\begin{center} 
\input{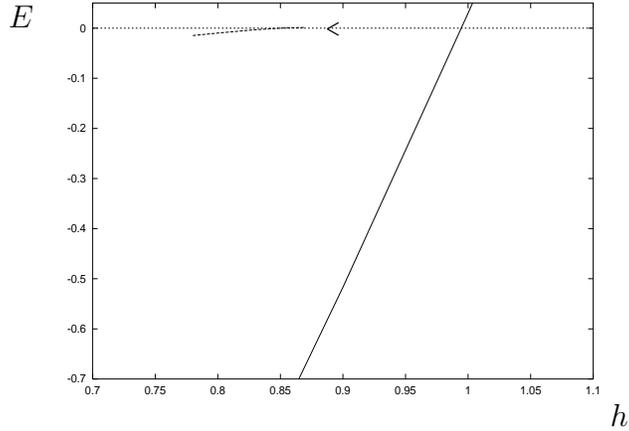}
\end{center}\vspace{-0.5cm}
\caption{The energy for $d=6.274$ $\kappa=0.35$}\label{ex2.196}
\end{figure}

\begin{figure}[ht]
\begin{center} 
\input{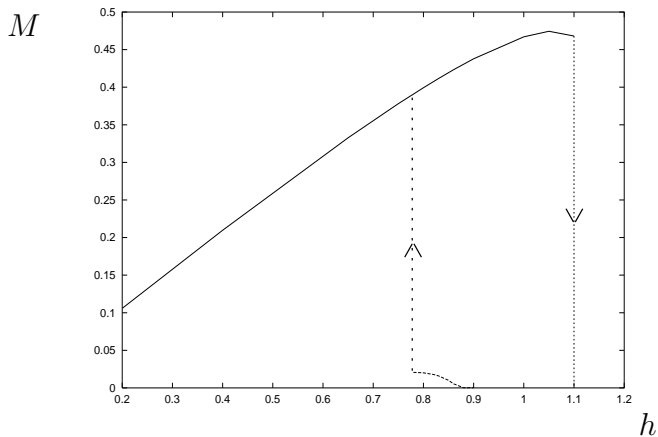}
\end{center}
\vspace{-0.5cm}
\caption{The magnetization for $d=6.274$ $\kappa=0.35$}
\label{mx2.196}
\end{figure}

\noindent
For parameters in this region, 
that is large domains and intermediate $\kappa$ (in a relative sense), 
a typical phase diagram 
is illustrated in Figure \ref{ux2.196} with the energy in Figure \ref{ex2.196}
and the magnetization in Figure \ref{mx2.196}. Three solution branches are
shown which represent the normal solution, the superconducting solution 
(solid line) and a solution with a single vortex (dashed line). 
A profile for one of 
the vortex solutions is given in Figure \ref{fig.01}.

The superconducting solution displays a hysteresis behavior as before, but
when increasing the field, it turns normal instead of going on the vortex branch as in Region 2. More precisely, as the field is increased, the superconducting solution loses its global stability, then its local stability and its drops
 to the normal branch when the super-heating field is reached. Conversely when decreasing the field from the normal state, the solution gets on the vortex branch, 
though it is only a local minimizer. So the transition when going down the field is of second order 
as can be seen on the magnetization curve (Figure \ref{mx2.196}), but when 
going up, there is 
a hysteresis.
 When decreasing the field further, the solution jumps to the superconducting branch.
 The vortex solution is never a global minimizer. 
In fact, our numerical experiments indicate that there are 
only locally stable vortex solutions.

We note that, 
for values of the Ginzburg-Landau 
parameter $\kappa$ in this region, 
the vortex state has not been frequently studied in the literature, except
for superconducting sample with extreme geometrical conditions such as
thin films, disks and rings. In the latter cases, the material displays
typical type-II behavior for all ranges of $\kappa$ as the Ginzburg-Landau
models can be simplified to allow an almost uniform penetration of the
magnetic field \cite{CDG}. However, the current 
study is done for three dimensional infinite cylinders and the simplified
models are not directly applicable. In fact, from the plot of the magnetic
field given in Figure \ref{fig.01}, we see that there is considerable
variation in the field strength over the cross section.

\hfill

\noindent
{\bf Curve $\kappa_3(d)$:}

\noindent
Let us call $H_c$ the thermodynamic critical field introduced by Ginzburg \cite{G}: the energy of the superconducting solution is equal to the energy of the normal solution at this field (in our nondimensionalization, 
it means that the energy of the superconducting solution is zero).
The curve $\kappa_3(d)$ corresponds to the situation 
where there is a  small amplitude vortex solution bifurcating from the normal solution exactly at $H_c$. 
One could hope to determine this curve mathematically.

 We notice that as $d$ tends to infinity, $\kappa_3(d)$ tends to a finite limit close to 0.4 which is less than $1/\sqrt 2$. Using (\ref{highd}), one can get that in the high $d$ limit, $\kappa$ is close to $\lambda_1 H_c$, where $\lambda_1$ is the first eigenvalue of the linearized problem (\ref{lin}) in an infinite domain and is close to 0.59.
 In Ginzburg's computations \cite{G} (see also \cite{T}), $H_c$ is exactly $1/\sqrt 2$ in the limit $d=\infty$. This gives $\lambda_1 /\sqrt 2$ as a limit for $\kappa_3(d)$ at infinity, which is close to $0.42$. A rigorous mathematical justification of this 
asymptotic behavior remains to be provided.

Let us recall that the critical value of $\kappa$ equal to $1/\sqrt 2$ that separates type I and type II superconductors was discovered by Ginzburg studying the bifurcation of the {\em superconducting } solution from the normal state for an infinite domain. Then Saint James and de Gennes discovered surface superconductivity, that is superconductivity is nucleated first in the surface and thus appears for higher fields than was calculated by Ginzburg.
 Here, we see that for $\kappa$ less than $\kappa_3(d)$, the vortex solution is no longer stable but is locally stable near the bifurcation.

\hfill

\noindent
{\bf Region 4: $\kappa <\kappa_4(d)$ and $d>d_1(\kappa)$.} 

\begin{figure}[ht]
\begin{center}
\input{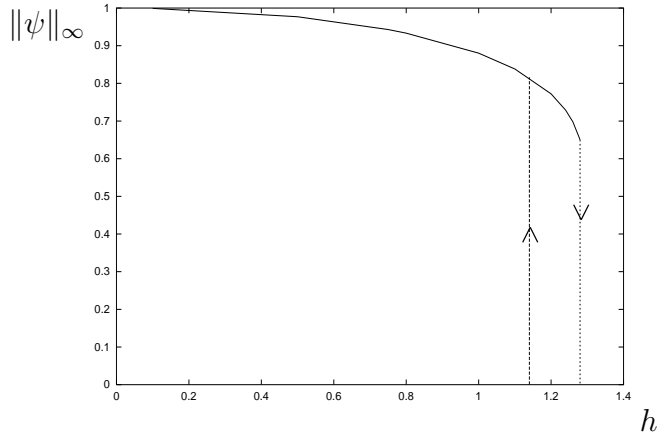}
\end{center}
\vspace{-0.5cm}
\caption{The bifurcation curve for $d=4$ $\kappa=0.3$}\label{ux1.2}
\end{figure}

\begin{figure}[ht]
\begin{center} 
\input{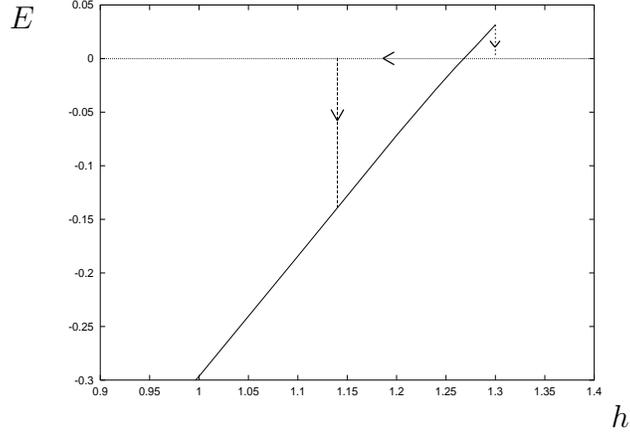}
\end{center}
\vspace{-0.5cm}
\caption{The energy for $d=4$ $\kappa=0.3$}\label{ex1.2}
\end{figure}

\begin{figure}[ht]
\begin{center} 
\input{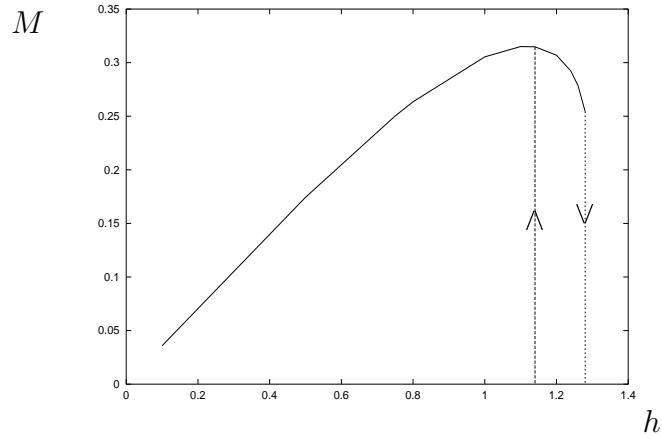}
\end{center}
\caption{The magnetization for $d=4$ $\kappa=0.3$}
\label{mx1.2}
\end{figure}

\noindent
For parameters in this region, that is $\kappa$ small but domains large enough, the typical bifurcation diagram 
is illustrated in Figure \ref{ux1.2} with the energy in Figure \ref{ex1.2}
and the magnetization in Figure \ref{mx1.2}.

There are superconducting solutions displaying a 
hysteresis phenomenon and no locally stable vortices. 
The superconducting solution is not always the global minimizer, 
but when increasing the field, the sample remains superconducting 
until reaching a super-heating field $h^*$, where the 
solution becomes normal with a discontinuous transition. Similarly, 
when decreasing the field, the sample stays normal until the field 
 $h_*$ which is less than $h^*$, where it turns superconducting by 
a discontinuous transition. Mathematically, we believe that there is 
a range of fields, between $h_*$ and $h^*$ where there are multiple 
superconducting solutions. This is in analogy with what happens for the 
one dimensional case. For more rigorous analysis of the one dimensional 
models, we refer to \cite{AT1}. 

Next, we also note that there is no locally stable vortex solution
in the region. 	
It is well-known that asymptotically for small $\kappa$, the vortex solution
is not energetically favorable, and 
the material  belongs to the typical
type-I regime  \cite{DG,T} where the phase transition is characterized by
the superconducting/normal interface rather than the vortex state.
When varying the field, the superconducting or normal solutions will not 
turn into a one vortex solution. 
However, if we do continuation from a vortex solution 
for a bigger value of $\kappa$ (in Region 3), and decrease $\kappa$,
we can still find existence of solutions with
vortices when we reduce $\kappa$ to values in Region 4 despite of 
the instability of vortex solutions.  
For even smaller $\kappa$, continuations in $\kappa$ or in other parameters
from the vortex solutions fails to produce any
new vortex solutions. We believe that when decreasing $\kappa$, the vortex solution first loses its stability near the normal solution (on the curve $\kappa_4(d)$, but it remains locally stable for $\|\psi\|_\infty$ a little higher in the branch. For very small $\kappa$, (especially less than $C /d$),
 we believe that 
there is no vortex solution at all. This has been proved in \cite{BPT} in the case of a disc. It is an  open problem to prove that for fixed $d$ and for $\kappa$ small enough, vortex solutions do not exist.

\hfill

\noindent
{\bf Curve $\kappa_1(d)$:}

\noindent
If $\kappa$ is fixed below 0.7 and $d$ is increased, then the curve $d_1(\kappa)$ is crossed. It remains establish the mathematical existence of this curve. In the particular case where $\kappa$ is very small, the order parameter $\psi$ is almost constant, there are no vortices in the domain so that  system (\ref{GL})
 simplifies to
$$\Delta A=|\psi |^2 A, \ \hbox{in}\  \Omega\quad \curl A=1 \ \hbox{on}\ \partial \Omega$$
where $|\psi |$ is a constant that depends on $h_0$. The boundary condition for $\psi$ yields $\int |\psi | (|\psi |^2+h_0^2A^2-1)=0$, which is the equation of the bifurcation curve.  One has to find the critical $d$ for which the curve $|\psi |(h_0)$ changes direction of bifurcation near $|\psi |=0$, so that the bifurcation goes from stable for small $d$ to unstable for larger $d$. Another way to study this curve is to make the bifurcation analysis near the normal state, described in the analysis of $\kappa_2(d)$, which yields to (\ref{lin}). Then one would need to take this development to higher order to get the sign of energy of the bifurcated branch. This sign changes on $\kappa_1(d)$. The fact that the bifurcation from the normal state is unstable for large $d$ has not been studied.

\hfill

\noindent
{\bf Curve $\kappa_4(d)$:}

\noindent
Another open problem is to determine the behavior of $\kappa_4(d)$ as $d$ tends to infinity. We expect it to be of the order $C/d$ for some constant $C$. We believe that the analysis that we have explained for the curve $\kappa_2(d)$ is the same here, that is on $\kappa_4(d)$ as well the eigenvalue has two eigenfunctions.  The same analysis of the linearized problem needs to be performed. The difference with $\kappa_2(d)$ is that the eigenfunction with no vortex is stable on $\kappa_2(d)$ and unstable on $\kappa_4(d)$.

\hfill

\noindent
{\bf The point of intersection}

\noindent
Note that all curves $\kappa_i(d)$ intersect at the same point. Indeed the point of intersection of $\kappa_2(d)$ and $\kappa_4(d)$ has an eigenvalue with two eigenfunctions, one of which (the one without vortices) changes stability. Hence this point also belongs to $\kappa_1(d)$ since on $\kappa_1(d)$ the stability of the solution without vortex changes. Finally this point belongs to $\kappa_3(d)$ since the energy of the bifurcated solution is zero for both eigenfunctions, in particular for the vortex solution. We want to point out that a similar analysis for the intersection of these curves has been performed in \cite{AC2} in the one dimensional setting and it yields to a solvability condition of fourth order at the point of intersection.

\section{Conclusion}
We have obtained very detailed bifurcation diagrams for 
the Ginzburg-Landau model
of a two dimensional cross section of a three dimensional superconducting
cylinder when the applied field is along the direction of the axis.
Detailed analysis are provided for the solution branches in the
different regions of the $(\kappa,d)$ plane. What distinguishes the different regions
 are features like the existence 
(or the lack of existence) of
vortex solutions, the global and local stability of solutions, and the
hysteresis phenomena. 
 We note that the analysis includes regions
with small and intermediate values of $\kappa$ 
which is not often featured in the existing studies,
as most of the works in the literature focus on the vortex state
which appears for larger values of $\kappa$ or for thin films.
Though we have computed the bifurcation diagrams
with rectangular cross sections, they are very representative of the
general cases. Naturally, for small samples, the geometric conditions
may have a stronger effect on leading to detailed alternations to the 
bifurcation curves. A change of topology, such as rings or shells, may
present other complications \cite{BPV} but we expect  similar bifurcation
diagrams remain valid. In addition, comparisons with existing 
theories have been made in the paper. 
Some remaining questions have also been raised.

\hfill

\noindent
{\bf Acknowledgments:} This work was made possible thanks to the joint 
research project France HongKong (Procore) grant.

\end{document}